\documentclass[%
 aip,
rsi,%
 amsmath,amssymb,
 reprint,%
]{revtex4-1}

\usepackage{graphicx}
\usepackage{dcolumn}
\usepackage{bm}
 \usepackage{url}
\begin{document}

\preprint{AIP/123-QED}

\title[Conducting lines]{Conducting lines - drawable conductors as a source of active learning activities}
\author{Petar Damjanovi\'c}%
\affiliation{Užice Grammar School, Trg Svetog Save 6, 31000 Užice, Serbia}%
\author{Vladimir Velji\'c}
\affiliation{Scientific Computing Laboratory, 
Center for the Study of Complex Systems,\\
Institute of Physics Belgrade, 
University of Belgrade, \\
Pregrevica 118, 11080 Belgrade, Serbia}%
\author{Aleksandra Alori\'c}
\affiliation{Scientific Computing Laboratory, 
Center for the Study of Complex Systems,\\
Institute of Physics Belgrade, 
University of Belgrade, \\
Pregrevica 118, 11080 Belgrade, Serbia}%

\date{\today}

\begin{abstract}
Drawable electronics attract the attention of both educators and innovative circuit engineers due to their affordability and simplicity. 
This paper focuses on active learning activities related to conducting lines and presents extensible laboratory projects suitable for students of all levels - from late primary school pupils to early undergraduate students.
We build on ideas of Pouillet's law classroom demonstrations with pencil drawings on paper and continue with a line, circle and grid drawings and appropriate resistance measurements. These are used as methodological examples of drawable circuits suitable for learning about resistors in series, parallel and mixed combinations. Particularly, grid drawings, e.g. mixed circuits, offer a variety of exercises that can connect analytical and experimental approaches and instigate discussions about assessments of uncertainty. Along the way, we propose ideas for extracurricular projects students can undertake to further their experimental knowledge and practise the scientific method. 

\end{abstract}

\maketitle

\section{\label{sec:Introduction}Introduction}

The affordability of paper and the simplicity with which customizable electric circuits can be drawn using graphite pencils and other conducting inks, make it appealing for drawable paper electronics~\cite{tobjork2011paper,kurra2013field, kurra2013pencil,costa2018hand}, foldable electronics~\cite{liu2017rollerball},  pencil–paper-based on-skin electronics~\cite{xu2020pencil}. For the same reasons, pencil and paper can be utilized in any classroom to effectively demonstrate key messages of basic and more advanced electronics curriculum~\cite{inyeo2018pencil,derman1999pencil,marshall2003pencil}.

This paper focuses on the educational aspect of conducting lines and the ways in which this simple experimental set-up can be used to illustrate physics phenomena typically taught in theoretical and mathematically technical terms.
The topic of electrical conductivity is commonly present in the early high school physics (or STEM) curriculum\footnote{In Serbian curriculum, the topic is already introduced in the eight (final) grade of the primary school (students' age of 15) and then discussed more deeply in secondary school, usually in second or third grade depending on the school's profile.}, and as such shows great potential for a variety of experimental approaches and hypothesis testing of interest for students of different age and knowledge level. 
Building on ideas for classroom demonstrations~\cite{kamata2012hand,chiaverina2014exploring}, we show that a similarly simple experimental set-up can be used to introduce students to systematic experimentation and the scientific method.
Furthermore, these topics can motivate a plethora of ideas for independent research regardless of the availability of extraordinary research facilities. 

The research topic and the student-led research approach have been greatly inspired by the International Young Physicist's Tournament (IYPT)\cite{IYPT1,IYPT2,plesch2020iypt}. 
The IYPT is a competition that fosters scientific investigation and debate among high school students for more than thirty years.
Every year 17 open-ended problems are chosen by the International Organizing Committee and students start their individual and team investigations that culminate a year later with a series of debates, e.g. 'physics fights' when students challenge each others' scientific results. One problem in the 2020 cycle of the IYPT initiated the study presented here: ``A line drawn with a pencil on paper can be electrically conducting. Investigate the characteristics of the conducting line.'' \cite{IYPTproblems2020}
As with many other IYPT problems, this simple formulation of the phenomena invites students to do in-depth literature research, come up with various hypotheses and devise an adequate experimental setup to test them. 

This makes the IYPT problem archive~\cite{IYPTarchive} an excellent database for student-led project work~\cite{planinsic2009iypt}. 
However, devising classroom interventions inspired by these problems that engage a wider group of students (not only those already interested in physics) is not an easy task. 
That is why in this paper we provide detailed ideas and instructions for classroom activities ranging from a simple class demonstration (\ref{sec:classroomdemo}), over ideas for physics laboratory (or physics practicum, \ref{sec:lab}) to project ideas (\ref{sec:projects}) that can be used to foster Investigative Science Learning Environment (ISLE)~\cite{etkina2015millikan} in classrooms. 

\section{\label{sec:classroomdemo}Classroom demonstration}

\begin{figure}[t!]
    \centering
    \includegraphics[width=0.45\textwidth]{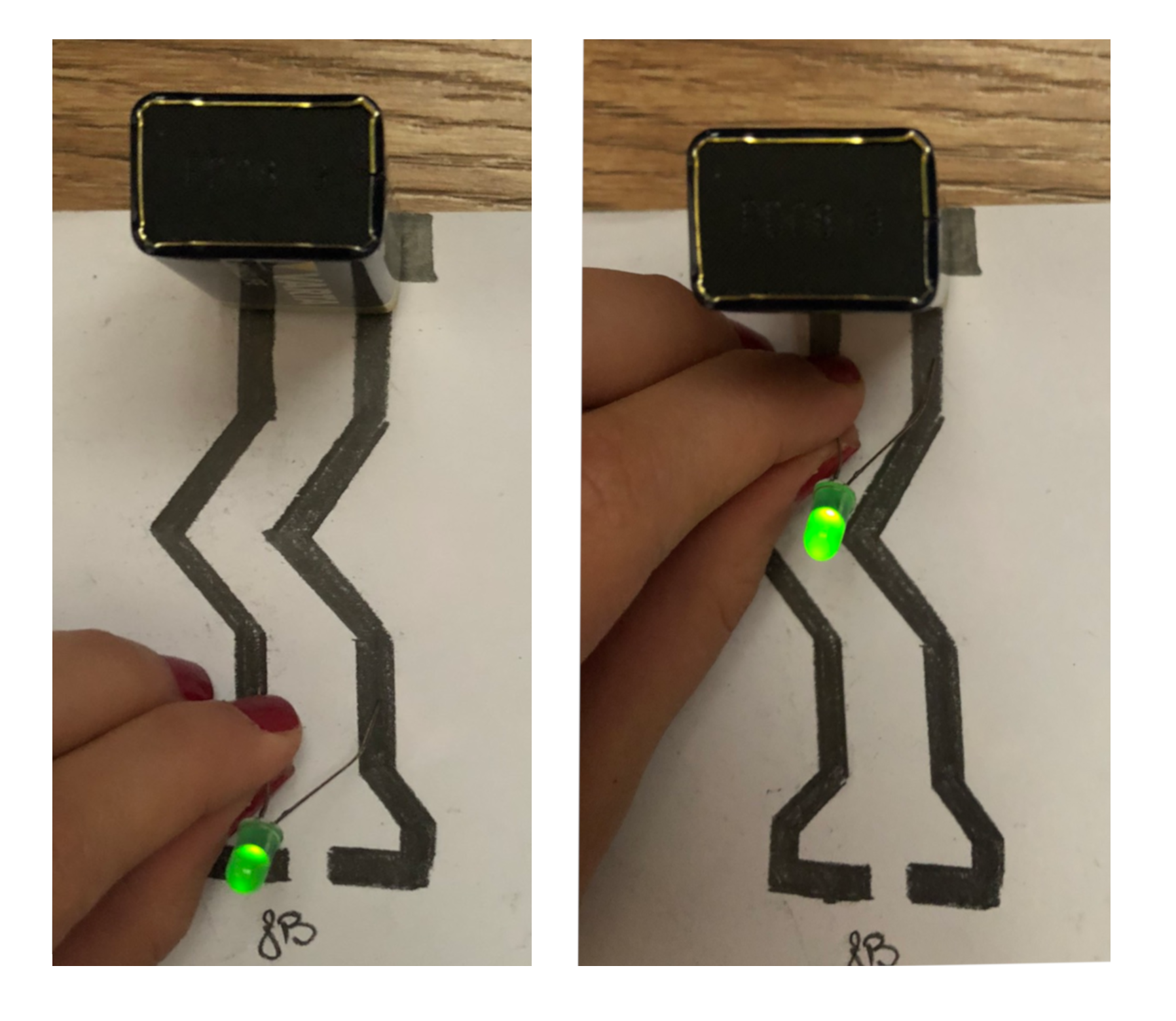}
    \caption{A simple and safe home experiment showing that in a circuit with a shorter line the diode lights brighter.
    This is an illustration from a student's submission during IYPT 2019 competition in Serbia, published with the permission of the author Una Ja\'cimovi\'c.}
    \label{fig:classroomdemo}
\end{figure}

Despite the popularity of conducting lines in physics education literature~\cite{chiaverina2014exploring,kuccukozer2015teaching,kamata2012hand}, science exhibitions~\cite{museum} and many commercial kits~\cite{circuitscribe}, the topic of electronic circuits and resistance is often thought of in mathematically technical terms through Ohm's law or Pouillet's law.
And although many students acquire these equations and use them readily to solve exam problems, this does not imply comprehension of concepts used~\cite{ibrahim2012representational}.
This is why we suggest that every lecture on electrical resistance starts with a simple demonstration~\cite{kamata2012hand} shown in Figure~\ref{fig:classroomdemo}. This demonstration needs only a single battery (9 V battery was used), a LED, paper and pencil (8B pencil was used), but it can help sparkle discussion among students and generate questions, e.g. given fixed input voltage (battery) is the light brighter for a shorter or longer line. What about line thickness? What if the line is closed? 

Alternatively, pencil leads (instead of drawn lines) can be used for demonstration and students can compare LED brightness for the conductors of the same length~\cite{inyeo2018pencil,kuccukozer2015teaching}. As discussed previously~\cite{kuccukozer2015teaching, woolf1996graphite}, pencil leads can be joined together (in series or in parallel), or lines of different lengths and width can be drawn to introduce key conductor's shape properties that regulate its resistance. Thus, students can grasp Pouillet's law inductively, observing the demonstrations
\begin{equation}
R \propto\frac{l}{A} \, ,
\end{equation}
where $R$ denotes electrical resistance, $l$ the resistor's length, $A$ its cross-sectional area. Furthermore, using pencils with different graphite content (e.g. HB scale, higher B numbers correspond to higher graphite to clay ratios~\cite{sousa2000observational}), students can deduce that the constant of proportionality (resistivity $\rho$) is material dependent while using dimensional analysis, they can deduce the constant's units. 
Only then, would we suggest that the educators take the path of introducing the full theoretical framework as described here~\cite{grisales2016preparation}. 

Building on these demonstrations and the discussions it has inspired among students, educators can make interdisciplinary connections with chemistry diving deeper into explanations of graphite structure~\cite{adorno2017amazing}.

\section{\label{sec:lab}Experiments suitable for individual or group based activities}

Expanding on students' discussions initiated with simple demonstrations presented in the previous section, the next step is to devise experiments to test students' generated hypotheses (e.g. dependencies formulated by Pouillet's law) in a systematic manner. 

In this section, we will overview three groups of experiments - serial, parallel and mixed circuits made of conducting lines. Materials needed include pencils of different graphite content (we used pencils with high graphite content: 5B to 8B), matte paper (most of the printing papers without a glossy coat are suitable, we used standard $80\:\rm g/m^2$ printing paper) and a digital ohmmeter. Due to the sensitivity of the drawn conductors to the pressure applied in drawing, all the experimental results presented here are drawn by the same person, following the same protocol - line of the desired length and fixed width is traced six times, aiming to apply the same pressure while drawing. Unless otherwise stated the line length was $12\:\rm cm$, while the width was  $1\:\rm mm$. The resistance of a given conductor was always measured five times placing probes at a fixed distance in a way that maximises the probe and line contact. In the plots, we show mean resistance with error bars denoting maximal measurement deviation from the mean. To exclude possible effects of the paper's roughness anisotropy, all the lines were drawn along the same paper's direction.

\subsection{\label{sec:series}Resistor lines - a series circuit}

The investigation of graphite lead resistance as a function of its length~\cite{inyeo2018pencil,kuccukozer2015teaching}, and similarly for the graphite drawn lines~\cite{chiaverina2014exploring,grisales2016preparation,derman1999pencil,kamata2012hand} is a simple introductory laboratory work students can do to test the previously generated hypothesis that resistance is proportional to resistor's length. Here we suggest that besides reinforcing Poulette's law in students' minds, the same exercise can be used to think about the equivalent resistance of resistors connected in series.

\begin{figure}[b!]
    \centering
    \includegraphics[width=0.45\textwidth]{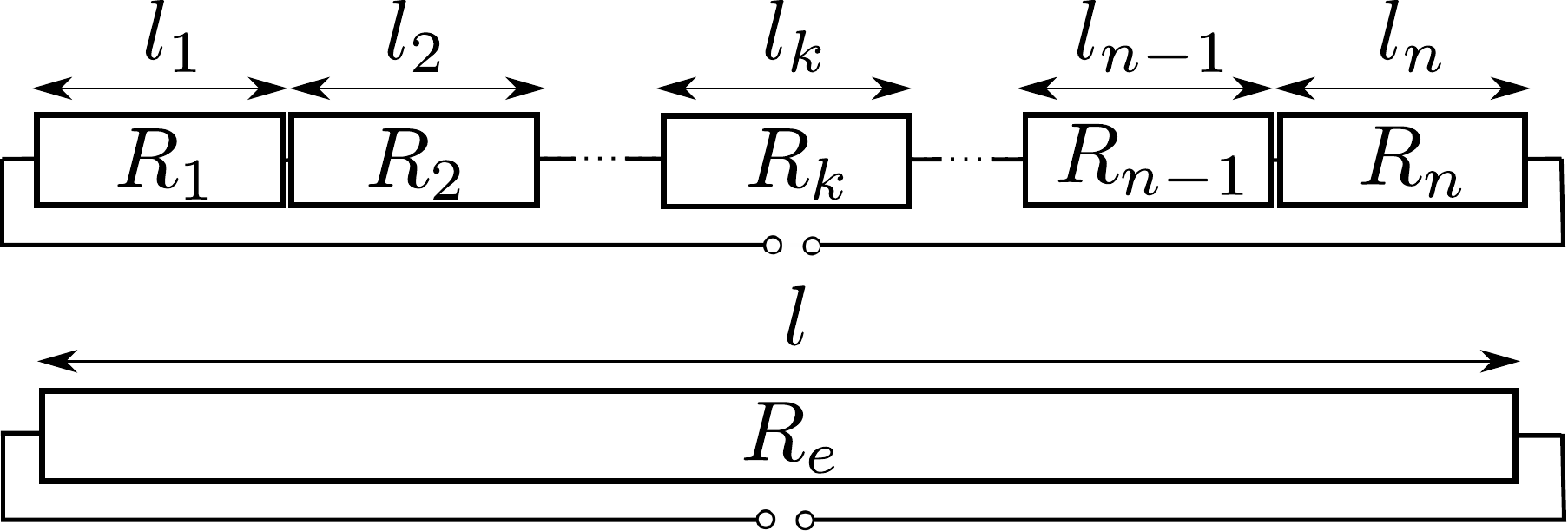}
    \caption{Illustration of the series circuit.}
    \label{fig:serial_resistivity_theory}
\end{figure} 

\begin{figure*}[t!]
\centering
\includegraphics[width=\textwidth]{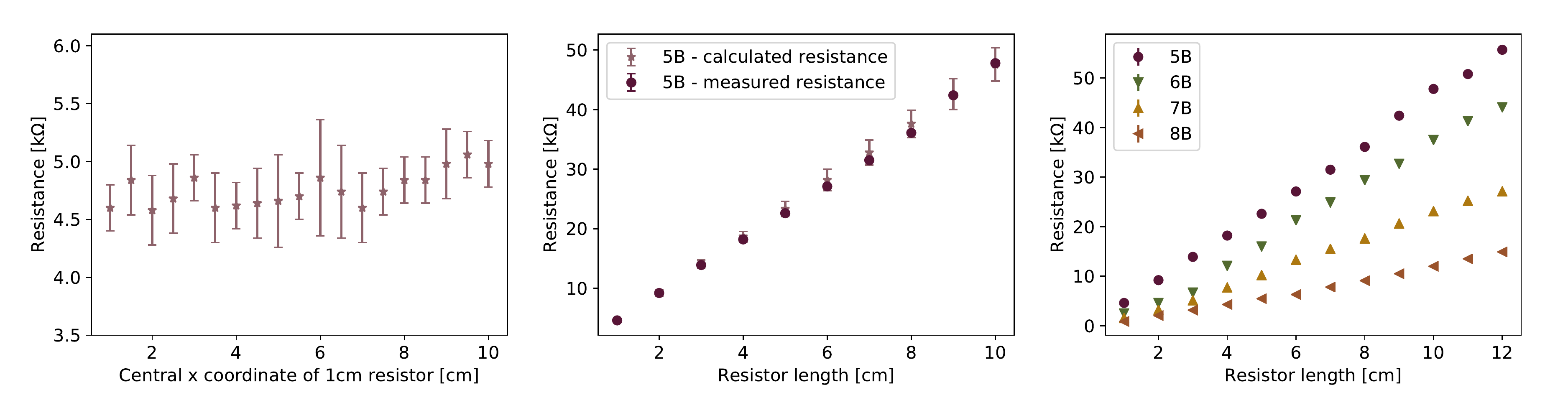}
\caption{Left: Resistance of centimetre-long line segments within the longer graphite line (5B pencil was used). Middle: Line resistance as a function of its length: measurement and serial resistance calculation comparison (5B pencil). Right: Line resistance as a function of its length for different pencil types (5-8B pencils).}
\label{fig:serial_resistivity}
\end{figure*} 

The students start by drawing a line resistor of fixed width and length. In the experiments shown in Fig~\ref{fig:serial_resistivity} the lines were $12\:\rm mm$ long and $1\:\rm mm$ wide. The first measurements students can perform are resistance measurements of the short line segments (in Fig.~\ref{fig:serial_resistivity} (Left) we show results for $1\:\rm cm$). Through this measurement, students effectively investigate the uniformity of the drawn line. As demonstrated in Fig.~\ref{fig:serial_resistivity} (Left), the segments' resistance variation is within the margins of error, confirming our assumption that the cross-section of graphite line, as well as its resistivity, can be treated as constant along the drawn line. With the measurements of individual resistances of line segments $l_i=1\:\rm cm$, students can proceed to calculate resistance for any length $l$ realising that the same way the total length can be broken into segments
\begin{equation}
    l = l_1+l_2+ \dots +l_k+ \dots + l_{n-1}+l_n \, .
\end{equation}
As illustrated in Fig.~\ref{fig:serial_resistivity_theory}, students can compare measured resistance of any given line length, but also calculate it using individual measurements of segments' resistances
\begin{equation}
    R_e = R_1+R_2+ \dots +R_k+ \dots + R_{n-1}+R_n \, .
\end{equation}
Results of one such comparison is shown in Fig.~\ref{fig:serial_resistivity} (Middle). This simple lab exercise where a single line's resistance is directly measured and calculated based on multiple measurements can lead to a fruitful discussion about measurement errors - the errors always increase when multiple measurements are combined. This point is clearly demonstrated by increasing margins of error in the Middle panel of Fig.~\ref{fig:serial_resistivity}.

Finally, we suggest that students proceed to measure how resistance varies with the length for pencils with different graphite content (in Fig.~\ref{fig:serial_resistivity} (Right) we use pencils 5B to 8B). The key message is an experimental confirmation of the demonstration induced hypothesis that resistance increases with the resistor's length. Also, students can realize that resistance increases faster for pencils with lower graphite content (5B pencil's slope is higher than 8B pencil's slope), which can help students deduce that the non-graphite content in pencils is a non-conducting material, usually clay.

The results of this exercise students can compare with results presented previously~\cite{inyeo2018pencil,grisales2016preparation,chiaverina2014exploring}, but also calculate resistivity per line's thickness~\cite{inyeo2018pencil} from experimental line slopes. Additionally, comparing the resistivity of a drawn line with the one of the pencil lead where all the dimensions can be measured, students can try to estimate the thickness of the drawn line~\cite{woolf1996graphite}.

\subsection{Resistor circles - a parallel circuit\label{sec:parallel}}

\begin{figure}[b!]
    \centering
    \includegraphics[width=7cm]{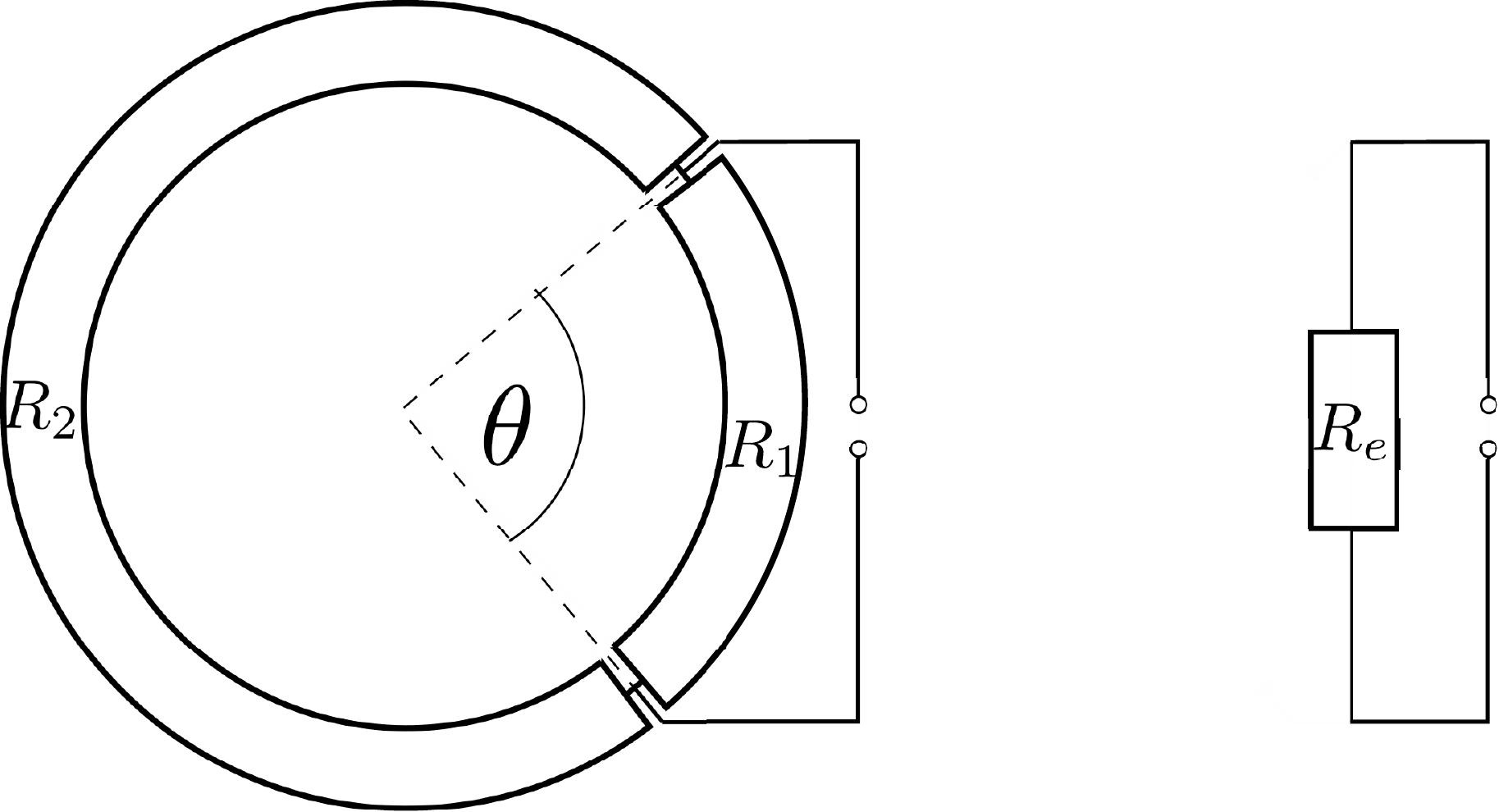}
    \caption{Illustration of the parallel circuit.}
    \label{fig:paralel_resistivity_theory}
\end{figure} 

\begin{figure*}
    \centering
    \includegraphics[width=\textwidth]{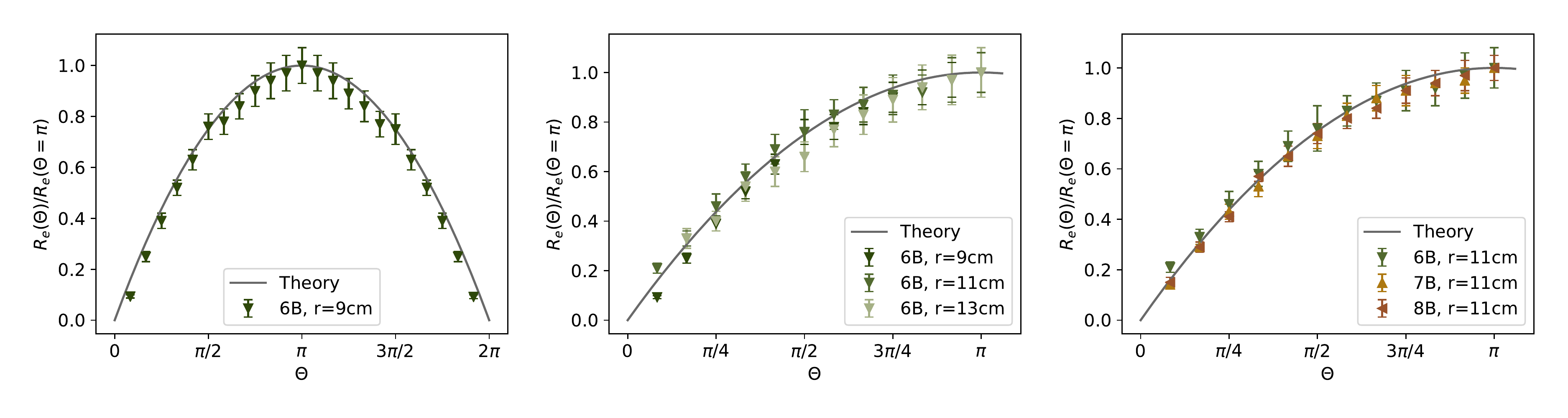}
    \caption{Left: The equivalent resistance of a circular conductor as a function of angular separation of the probes. Middle: The equivalent resistances of circles with different radius. Right: The equivalent resistances of circles drawn with different pencils.}
    \label{fig:paralel_resistivity}
\end{figure*} 

Contrary to the graphite series circuit demonstrations and particularly, resistance as a function of length, the graphite parallel circuits classroom explorations are less common. Kamata and Abe~\cite{kamata2012hand} use parallel conducting lines circuits with LED to show that the same resistor lengths connected in parallel result in brighter LED. 
Other authors~\cite{chiaverina2014exploring,woolf1996graphite,marshall2003pencil,kuccukozer2015teaching} address parallel circuits with proposed activities such as drawing parallel lines and connecting them in a circuit, for detailed instruction see Derman and Goykadosh~\cite{derman1999pencil}.
Here instead, we propose a series of systematic measurements that can be done on a resistor drawn in form of a circle.

A circle-shaped resistor of radius $r$, cross-section area $A$ and resistivity $\rho$ divided in two segments by probes placed at angular distance $\theta$ (see Fig.~\ref{fig:paralel_resistivity_theory}) exemplify a parallel circuit of two resistors. The reciprocal of the equivalent resistance of resistors connected in parallel equals the sum of reciprocals of each resistor in the circuit 
\begin{equation}
    \frac{1}{R_e} = \frac{1}{R_1}+\frac{1}{R_2} \, .
\end{equation}
Both resistors' resistances can be calculated (as discussed in previous section) by using the appropriate arcs' lengths, e.g. $l_1=r \theta$ and $l_2=r (2\pi-\theta)$. Thus the equivalent resistance of a circular resistor measured with the probes separated by angle $\theta$ reads
 \begin{equation}
    R_e = \frac{\rho r}{A} \frac{\theta(2\pi-\theta)}{2\pi} \, .
\end{equation}
Getting this $R_e(\theta)$ dependence represents an interesting mathematical exercise,  but at the same time excited  experimental challenge since measured dependency between resistance and angle is not linear as students often expect~\cite{van2004remedying}.

We propose that students do this experiment in pairs or groups focusing on different possible parameters they might want to explore, e.g. circles of different radius, different number of line tracings over the circle, different pencil grade, etc. To compare different results, we propose that the equivalent resistance is normalized with $R_e(\theta=\pi)$, which yields to
\begin{equation}
    \frac{R_e(\theta)}{R_e(\theta=\pi)} =\frac{\theta(2\pi-\theta)}{\pi^2} \, .
\label{eq:universalcirc}
\end{equation}

As the normalized equation is just $\theta$ dependent it represents a universal result. In other words, this function is independent of other line's parameters and thus offers an excellent opportunity for multiple student groups' results to be compared together against the same theoretical line.  
Since this $\theta$-dependence has the shape of an inverted parabola, it is worth emphasizing that it reaches a maximum value of 1 when $\theta = \pi$, which corresponds to the maximum possible equivalent resistance of this configuration. Analogously, the minimum value of this function is 0 when $\theta = 0$ or $\theta = 2\pi$, which corresponds to the minimum theoretical equivalent resistance of this configuration.

In Figure~\ref{fig:paralel_resistivity} we present measurements from several experiments against the theoretically obtained function (Eq.~(\ref{eq:universalcirc})) with overall excellent agreement. The circles of stated radius were drawn by outlining different cylinders. One of the probes was fixed, while the other was moved along the circular arc in steps of $\pi/12$.

In Fig.~\ref{fig:paralel_resistivity} (Left) we show full range of angles for the circle with radius $r= 9 \:\rm{cm}$ traced with 6B pencil and compare it with the theoretical expectation. For this combination of parameters, we demonstrate the full shape of the parabola to highlight symmetry and demonstrate uniformity of the drawn circle (as results for $\theta>\pi$ correspond to the analogous circuits to those $\theta<\pi$ if the angle is measured in the negative mathematical direction). On the other hand, the middle and right panels of the same figure depict the universality of the equivalent resistance for different choices of line parameters (radius and pencil type). Namely, normalization given by Eq.~(\ref{eq:universalcirc}) 
allows us to obtain magnificent quantitative agreement with experimental results as it collapses data onto a single curve.

\subsection{Resistor grids - a mixed circuit\label{sec:mixedcircuit}}
\begin{figure*}
    \centering
    \includegraphics[width=\textwidth]{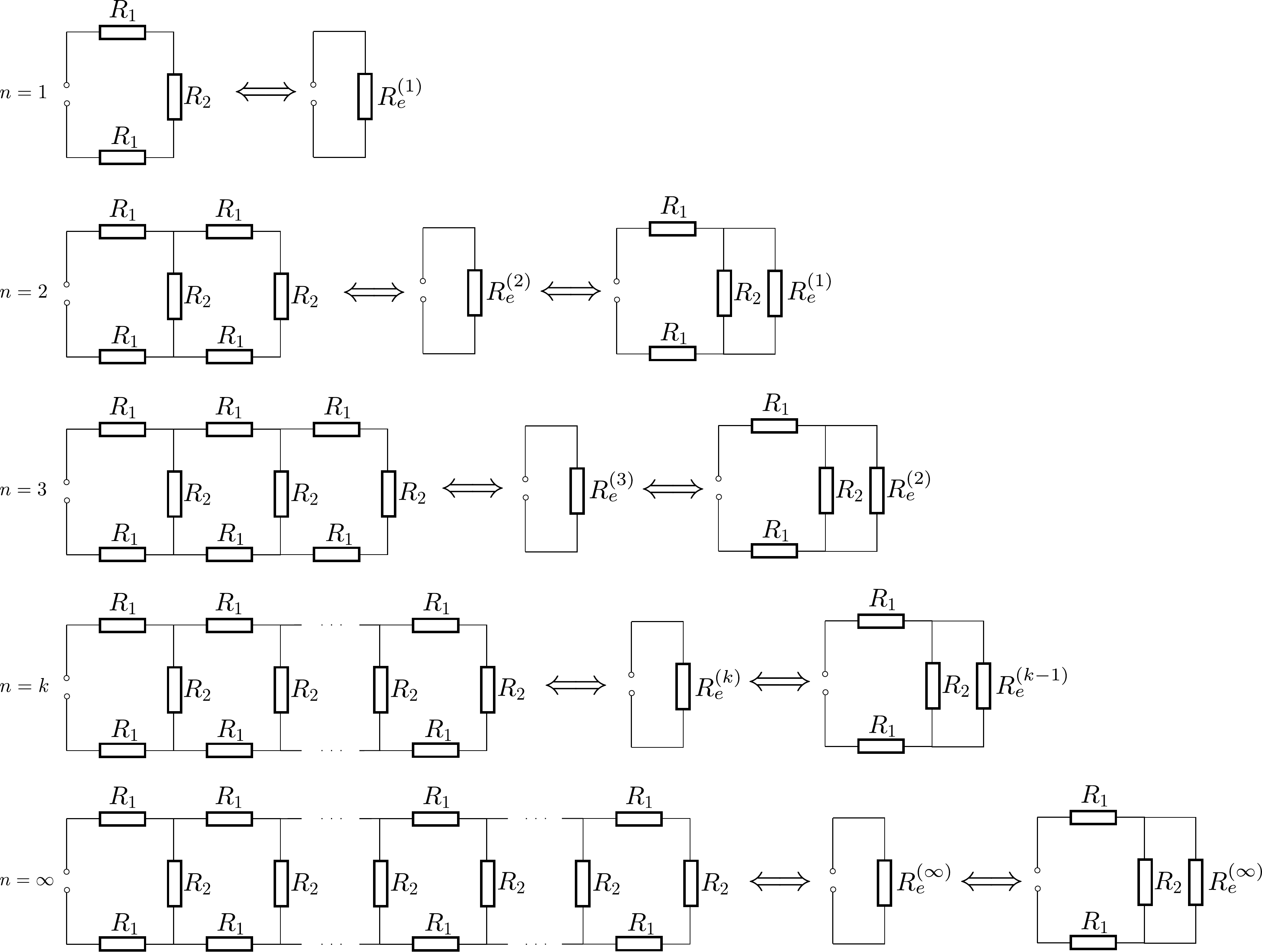}
    \caption{Illustration of a mixed circuit resistor grid.}
    \label{fig:mixed_circuit}
\end{figure*}

In the previous two sections, we have seen that resistors in series and parallel circuits, can be experimentally addressed with drawable resistors. Here we continue with mixed circuits. Particularly, we analyze mixed circuits that consist of a variable number of identical cells connected in parallel. The cells are constructed of two resistors $R_1$ and one resistor $R_2$ connected in series, as depicted in Fig.~\ref{fig:mixed_circuit}. This growing mixed circuit allows students to experimentally and theoretically investigate how the equivalent resistance changes with the number of cells in the grid.

When the circuit contains only one cell ($n=1$) the equivalent resistance $R_e^{(1)}$ is an equivalent resistance of resistors connected in series, given with Eq.~(\ref{eq:Re1}). For the case when $n=2$ the new cell can be replaced with a single resistor of resistance calculated in $n=1$ step - $R_e^{(1)}$, thus new equivalent resistance for $n=2$  can be rewritten as in Eq.~(\ref{eq:Re2}).  
Similarly, when the circuit consists of just $n=3$ cells the last two cells can be replaced with a single resistor of resistance $R_e^{(2)}$, so that equivalent resistance can be rewritten as in Eq.~(\ref{eq:Re3}). 
Namely, a specific relationship can be observed between the equivalent resistance of a network consisting of $n$ cells with the equivalent resistance of a network consisting of $n-1$ cells, and it is given by a recurrent relation in Eq.~(\ref{eq:Ren}). Having this relation we can calculate the equivalent resistance of such a mixed circuit for an arbitrary number of cells $n$.

\begin{align}
    R_e^{(1)}&=2R_1+R_2      \label{eq:Re1} \\
    R_e^{(2)}&=2R_1+\frac{R_2 (2R_1+R_2)}{R_2+ (2R_1+R_2)}  \nonumber \\ 
    &=2R_1+\frac{R_2 R_e^{(1)}}{R_2+R_e^{(1)}}  \label{eq:Re2} \\
    R_e^{(3)}&=2R_1+\frac{R_2 \left(2R_1+\frac{R_2 (2R_1+R_2)}{R_2+ (2R_1+R_2)}\right)}{R_2+ \left(2R_1+\frac{R_2 (2R_1+R_2)}{R_2+ (2R_1+R_2)}\right)} \nonumber \\ 
    &=2R_1+\frac{R_2 R_e^{(2)}}{R_2+R_e^{(2)}}  \label{eq:Re3} \\
   & \vdots \nonumber  \\
    R_e^{(n)}&=2R_1+\frac{R_2 R_e^{(n-1)}}{R_2+R_e^{(n-1)}} \label{eq:Ren}
\end{align}

\begin{figure}
    \centering
    \includegraphics[width=8cm]{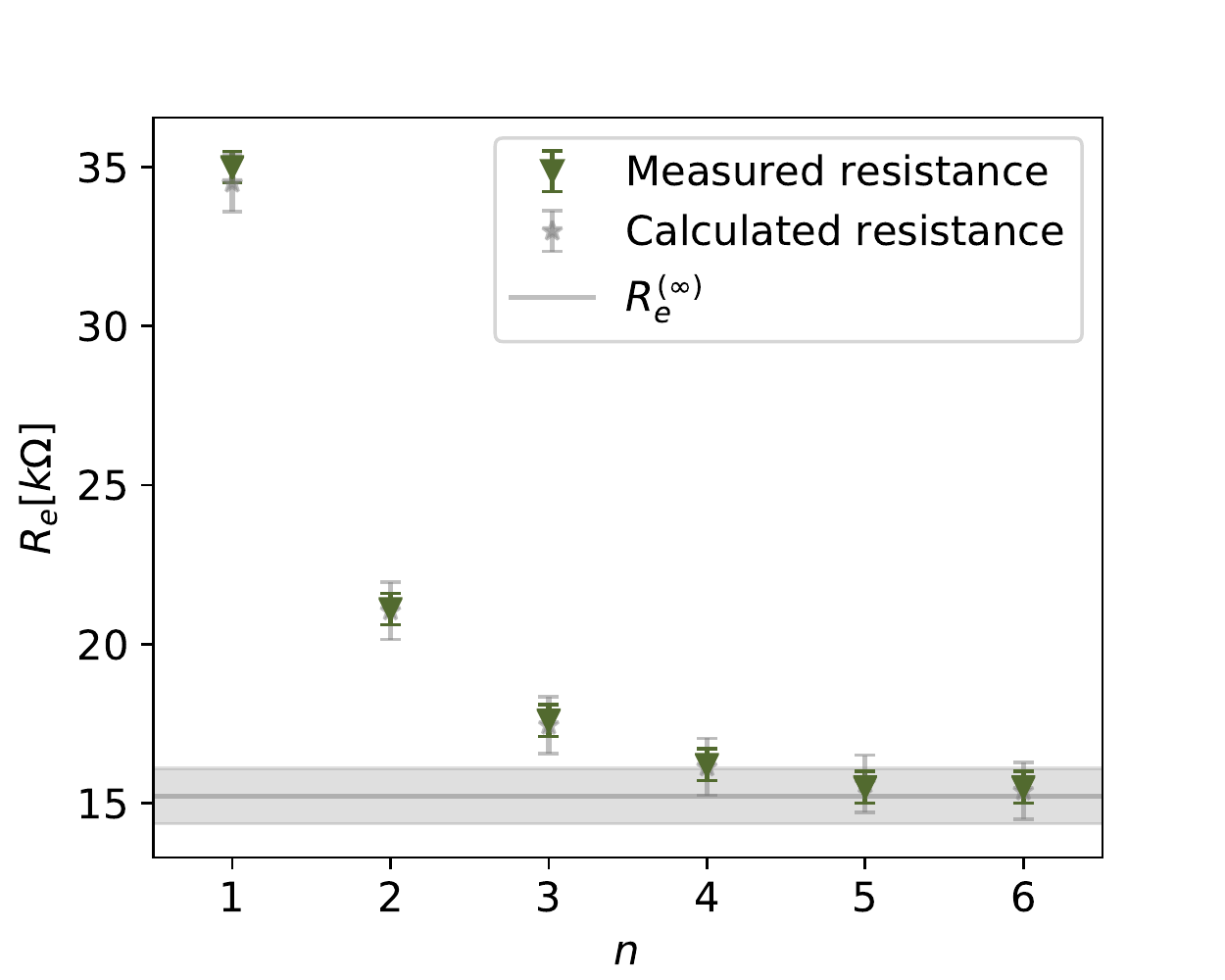}
    \caption{Comparison of measured and calculated equivalent resistance as a function of number of cells of mixed circuits depicted in Fig.~\ref{fig:mixed_circuit}.}
    \label{fig:mixedcircuit_experiment}
\end{figure}

As the electrical circuits as the ones shown in Fig.~\ref{fig:mixed_circuit} can be drawn, we can experimentally confirm results given by the Eqs.~(\ref{eq:Re1})-(\ref{eq:Ren}). In place of resistors denoted with $R_1$ using 6B pencil we draw lines of length $1\:\rm cm$, while for $R_2$ we draw lines of length $7\:\rm cm$, with all other parameters kept as previous (e.g. width, number of traces). Before connecting them in the circuits, we measure resistances of the lines $R_1=(2.6 \pm 0.2)\:\rm{k\Omega}$ and $R_2= (29.3 \pm 0.8) \:\rm{k\Omega}$.
We propose that initially only the first cell is drawn, while the sequence of resistors $R_2$, or the long parallel lines, is added equidistantly (one centimetre apart) so that only with tracing resistors $R_1$ the cells can be joined and added gradually as per Fig.~\ref{fig:mixed_circuit}. 
In Fig.~\ref{fig:mixedcircuit_experiment} we show the equivalent resistance of the mixed grid for each number of cells $n$, both measured and calculated using Eqs.~(\ref{eq:Re1})-(\ref{eq:Ren}).

We see a good agreement between the calculated and measured equivalent resistances across different cell numbers $n$. Additionally, we note that the equivalent resistance decreases with the number of cells. This is to be expected, as each new cell is added in parallel to the rest of the circuit, but this offers further space for discussion with students.
Since the equivalent resistance is decreasing with the number of cells, but it is also positive by the definition, it is implied that it must have a lower bound.
Theoretically, this bound can be calculated in the limit of infinite number of cells (see the last circuit in Fig.~\ref{fig:mixed_circuit}). The large $n$ limit allows us to exchange both $R_e^{(n)}$ and $R_e^{(n-1)}$ with $R_e^{(\infty)}$ in the Eq.~(\ref{eq:Ren}) leading to the following relation
\begin{equation}
    R_e^{(\infty)}=2R_1+\frac{R_2 R_e^{(\infty)}}{R_2+R_e^{(\infty)}} \, .
\end{equation}
The last equation is a quadratic equation which can be solved analytically and its solution is
\begin{equation}
    R_e^{(\infty)}=R_1+\sqrt{R_1^2+2R_1R_2} \, .
    \label{eq:Reinfty}
\end{equation}
Note that the second solution of the quadratic equation is rejected, since equivalent resistance has to be positive.
In Fig.~\ref{fig:mixedcircuit_experiment} horizontal line represents this theoretical lower bound, i.e. infinite $n$ circuit resistance. 
Calculating the large $n$ theoretical resistance value and comparing it with experimentally accessible number of circuit cells can help student grasp limiting values. In this particular example, we see in Fig.~\ref{fig:mixedcircuit_experiment} that already with $n=4$ cells, the limiting $n$ resistance value is reached (within error margins). So drawing circuits with $n\geq 5$ cells is practically indistinguishable from the point of effective resistance.

Both finite and $n\rightarrow \infty$ calculated resistances shown in Fig.~\ref{fig:mixedcircuit_experiment} have error margins and calculating them can be another theoretical exercise for students interested in error propagation. We provide these in the Appendix.

\section{\label{sec:projects}Independent research projects}

\begin{figure*}
    \centering
    \includegraphics[width=\textwidth]{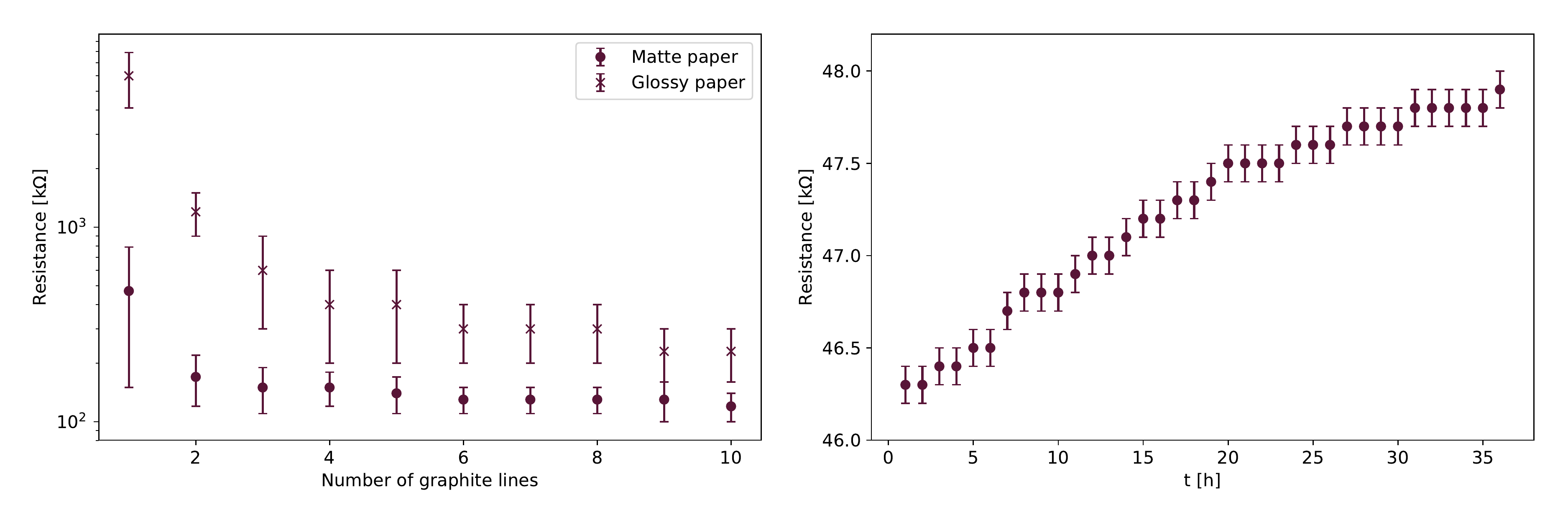}
    \caption{ Left: Line resistance as a function of the number of line traces for two types of paper denoted matte and glossy.
    Right: Line resistance as a function of time passed from the drawing. In both panels, 5B pencil was used for drawing $10 \:\rm cm$ long lines.}
    \label{fig:research_projects}
\end{figure*}

As it is evident from previous examples, this area of research offers a wide variety of ideas for independent research projects. Therefore, in this section, we propose several additional ideas that we have encountered.

Previously, we focused on the effects of line length as the main variable affecting the resistance and we aimed to keep the line thickness constant, e.g. by tracing the line the same number of times. In the exploratory analysis, we observed that a single traced line had too high resistance and was often out of our measurement reach, so we settled for retracing several times, in our case six, to ensure all required measurements will be within reach. To investigate systematically how the resistance depends on line thickness and with a simple experimental set-up described so far, we varied the number of tracings of a line with a fixed length. In Fig.~\ref{fig:research_projects} (Left) we show the measured resistances obtained by the same pencil lead (5B) and two different paper types. The line length is $10\:\rm cm$, while its other properties are kept as previously.
As expected resistance decreases when the line is traced multiple times. This was briefly explored previously~\cite{kamata2012hand,inyeo2018pencil}, on the smaller range of line tracing, reaching the same general conclusion. However, within our measured range, we see that the resistance has a lower bound and we hypothesize that there is a pencil content saturation and consequently limit in the line resistance, for the given length and width constraints of the line. Exploring how this limiting resistance depends on relevant paper and pencil parameters could become an engaging student's independent project.

Within the same experiment, we investigated the effects of different paper types
as it is clear that the conductivity of the drawn line depends also on the paper as that affects the amount of pencil trace left on the surface. In Fig.~\ref{fig:research_projects} (Left) we use two papers; one matte ($80\:\rm g/m^2$ printing paper used in experiments previously) and one glossy paper. As expected, the resistance of the line of the same length and the same number of line tracings (drawn by a pencil of the same lead type) was higher on the glossy paper. We hypothesize this is due to the lower amount of graphite content that is transferred to the paper.
Thus, we suggest further investigations of surface property influence on resistance by systematic variation of paper roughness, but also measuring line resistance when drawing on different surfaces (see for example resistance of lines drawn on MagicTM tape~\cite{derman1999pencil}, or a wider range of different surfaces~\cite{Shmavonyan2015graphite}).

Differently from the previous experiments, in Fig.~\ref{fig:research_projects} (Left) we show resistance measurements for three different lines made following the same protocol, denoted by mean and maximal deviations from it. This was done to estimate the differences that might occur among the lines when the lines are retraced as sometimes the graphite content might be removed with a new trace, or variations in pressure might become more visible. Due to this reason, the error bars are greater than elsewhere in the paper, where the key source of error is in the probe-line contact, but we note that the difference between line resistances drawn on different papers are statistically significant.

Finally, we report on a serendipitous discovery that could lead towards more interesting independent research activities. Within our reproducibility analysis, we observed that all resistance measurements a day later were offset by a constant value, while the paper with the drawn lines was kept under typical room conditions. This prompted a preliminary study whose results are shown in Fig.~\ref{fig:research_projects} (Right). Typically prepared resistor line of length $10 \:\rm cm$ was measured every hour for the total duration of $36$ hours. We observe that the resistance steadily increases within the first couple of hours, while after a day a change is much slower. Investigating whether the resistance reaches a plateau or keeps increasing at an even slower rate after the time window presented here could be a great practice in perseverance. To exclude the hypothesis that the increased resistance is a consequence of small graphite removal with every measurement, for this experiment we connected ohmmeter probes to the line and only read of resistance value hourly without further impact to the drawn resistor. We hypothesize that resistance change might be due to room humidity and absorption of water from the air. Confining the experimental set-up in an aquarium with a controlled humidity level (coupled with humidity measured by another excellent IYPT project - a hair hygrometer~\cite{IYPTarchive}) could lead to a stimulating research project.

Until further investigations that will detail the reasons behind line resistance changes in time, in this work we ensured that all the measurements are done within the first hour after the lines were drawn and we suggest similar practice, especially when aiming to investigate the effects of another parameter such as line length or width on the resistance. 

\section{\label{sec:Conclusions}Discussion and Conclusions}

In this paper, we have gathered a few simple, yet instructive and engaging experiments that rely on the typical students' supplies - pencil and paper - but instead of using them only for theoretical physics calculation, we suggest more active experimental application. A strong case for using pencil leads and traces at least for demonstrations of resistance was made by Kuccukozer~\cite{kuccukozer2015teaching} claiming that many teachers in Turkey omitted demonstration that resistance decays when resistor's length increases as it was not always evident to students when copper was used and students would even reach wrong conclusions. On the other hand with pencil trace and the simplest set-up as in Fig.~\ref{fig:classroomdemo} this result becomes evident. Other authors~\cite{kamata2012hand} even transform this simple demonstration to a form of black-box investigative exercise where students' goal is to deduce the line length or type or circuit drawn, given the LED brightness.

Besides classroom demonstration that can be effective and engaging, in this paper we covered a variety of ideas for extensive measurements that still do not require advanced equipment but allow practising experimental methods and scientific discussion between students. Due to the simplicity of the experimental set-up, many of the proposed exercises can be done individually or in groups. For example, in the activities presented in Sec.~\ref{sec:series}, all students can individually draw lines with different pencils or papers and perform uniformity measurements (e.g. produce plots like the one in Fig.~\ref{fig:serial_resistivity} (Left)). Students can then collectively discuss these results and decide which lines are the most suitable (uniform resistance) for further $R$ versus $l$ measurements which they can perform in pairs or groups and then aggregate them in a joint plot for discussion and conclusions. 
Students who worked using the same paper \& pencil combination can discuss differences in results introduced by different drawers and these discussions could lead to assessments of error margins and meaningfulness of different conductors (different paper/pencil) result comparison. 

Derman and Goykadosh~\cite{derman1999pencil} suggest measurements of circle arc resistances where linear dependence on the angle ($R\sim l \sim \theta$), similar to the one on the line length can be observed. Additionally, the authors report on measuring the value of $\pi$ by comparing the resistance of the arch and diameter. These measurements can be further discussed with parallel resistance measurements (discussed in \ref{sec:parallel}) as students can observe that drawn circle can be used to show both: a linear relationship between resistance and angle (when only the fragment of circle line is used) and quadratic relation (when the full circle is drawn). The resistance measurements discussed in Sec.~\ref{sec:lab} can inspire students and teachers to further connect the physics and mathematics curriculum by drawing different geometric shapes and using the methods discussed as ways to measure and calculate circumference or other geometric relations. This can be further complicated by encouraging students to create their extensions of resistor grids (e.g. triangular) for which they can measure and calculate equivalent resistance as described previously.

Besides ideas for further research discussed in Sec.~\ref{sec:projects} students can base their research on other types of conductive inks~\cite{grisales2016preparation,circuitscribe} and repeat any of the reported experiments or some of the proposed variations. Another interesting research direction next to exchanging paper for another type of surface could be a step towards the foldable electronics by either stretching, bending or folding the surface after drawing on it~\cite{inyeo2018pencil} and investigating how that affects measured resistances. Apart from resistors, drawings on the paper could be used to create capacitors~\cite{grisales2016preparation,salami2007capacitors} and more complicated circuits~\cite{kurra2013pencil,kurra2013field}, thus similar exercises could serve for further electronics classroom activities.

In line with previous work that advocates for measurement uncertainty as one of the key concepts students should grasp through laboratory experiments~\cite{trumper2002we}, various exercises we reviewed offer multiple learning opportunities in this direction. Starting with a discussion about device measurement errors (e.g. ohmmeter, ruler, protractor), continuing with measures of the agreement among the repetitive resistance measurement (in this case error mostly comes from probe-line contact) to uncertainty that is due to small variations in sample preparations (e.g. one centimetre segment resistances). Furthermore, students can discuss uncertainties from error propagation arising from multiple measurements (e.g. Fig.~\ref{fig:serial_resistivity} (Middle) and Fig.~\ref{fig:mixedcircuit_experiment}).

The simplicity and availability of the experimental set-up used in this paper allow students to familiarize themselves deeper with the topic of electric resistance, actively participate in designing apparatus and measurements and pose relevant research questions. Particularly, through different conducting lines and circuit types, we reviewed in the paper, students can grasp that straight line resistance linearly depends on its length and that circle resistance depends quadratically on angular probe separation. Finally, these exercises help students familiarize themselves with the scientific method of inquiry, processes of systematic measurements and the analysis of experimental results that are practical and applicable in a variety of different contexts besides the electric resistance of conducting lines.

\section{\label{sec:Acknowledgements}Acknowledgements}

A. A. acknowledges the funding provided by the Institute of Physics Belgrade, through the grant by the Ministry of Education, Science, and Technological Development of the Republic of Serbia. 
P. D. acknowledges valuable feedback and guidance from his physics teacher Cmiljka Vasovi\'c. 
All authors are grateful to the International Young Physicist Tournament, its Serbian community and participants for the inspiring atmosphere and scientific discussions which lead to this work, particularly, authors are grateful to Jelena Pajovi\'c. All authors thank Una Ja\'cimovi\'c, author of the Fig.~\ref{fig:classroomdemo}.

\nocite{*}
\bibliography{rad-bibliografija}

\appendix 
\section{Uncertainty analysis}

In line with previous work~\cite{trumper2002we} that singles out an estimate of uncertainty as one of the key features students should learn while doing laboratory experiments in order to grasp the scientific method, here we provide additional information on the error analysis for the calculated values presented in the paper. 

Within the paper, we focus most of the attention on the measurement errors and denote those appropriately within the plots. However, as discussed in the context of Fig.~\ref{fig:serial_resistivity}, some of the measurements provide learning opportunities to discuss propagation of errors where students can practically grasp the fact that an error of a summed variable is obtained by summing the errors
\begin{equation}
    \Delta R_e = \Delta R_1 + \Delta R_2+ \dots +\Delta R_n \, .
\end{equation}
The lecture on error propagation can be even further deepened in the Sec~\ref{sec:mixedcircuit} where students can assess errors not only for the directly measured resistances, but also for the equivalent resistances calculated using Eqs.~(\ref{eq:Re1}), (\ref{eq:Ren}), and (\ref{eq:Reinfty}). Here we provide absolute error relations that we used to calculate errors denoted in Fig.~\ref{fig:mixedcircuit_experiment} (for more information on error propagation see for example~\cite{ku1966notes}).
\begin{align}
    \Delta R_e^{(1)}=&2\Delta R_1+\Delta R_2 \, , \\
    \Delta R_e^{(n)}=&2\Delta R_1+\left(\frac{R_e^{(n-1)}}{R_2+R_e^{(n-1)}}\right)^2 \Delta R_2  \nonumber \\
    &+ \left(\frac{R_2}{R_2+R_e^{(n-1)}}\right)^2 \Delta R_e^{(n-1)}\label{eq:deltaRen}\, , \: n \geq 2 \, , \\
   \Delta R_e^{(\infty)}= &\left( 1+\frac{R_1+R_2}{\sqrt{R_1^2+2R_1R_2}}\right)\Delta R_1
    \nonumber \\
    &+\frac{R_1}{\sqrt{R_1^2+2R_1R_2}}\Delta R_2 \, . 
\end{align}

\end{document}